\documentclass[conference, 10pt]{IEEEtran}

\usepackage{cite}
\usepackage{graphicx}
\usepackage{textcomp}
\usepackage{xcolor}
\usepackage{booktabs} 

\usepackage[a4paper, total={184mm,239mm}]{geometry}
\def\BibTeX{{\rm B\kern-.05em{\sc i\kern-.025em b}\kern-.08em
    T\kern-.1667em\lower.7ex\hbox{E}\kern-.125emX}}
    
\usepackage{amsmath}
\usepackage{physics}
\usepackage{eucal}
\usepackage{xspace}
\usepackage{multirow}
\usepackage{algorithm}
\usepackage{algpseudocode}
\usepackage{enumitem}
\usepackage{caption}

\usepackage{comment}

\usepackage{soul,color}

 \usepackage{hyperref}
 \hypersetup{
     colorlinks=true,
     linkcolor=blue,
     filecolor=blue,
     citecolor=blue,      
     urlcolor=cyan,
     }
\usepackage{cleveref}

\newcommand{\name}{qGDP\xspace}
\newcommand{\crossing}{$\bar{\text{X}}$}

\newcommand{\x}{$\times$\xspace}

\newcommand{\Q}{$t_q$}
\newcommand{\C}{$t_e$}

\BeforeBeginEnvironment{equation}{\begingroup\footnotesize\vspace{-5pt}}

\AfterEndEnvironment{equation}
{\endgroup}

\author{
Junyao Zhang\textsuperscript{1}, 
Guanglei Zhou\textsuperscript{1}, 
Feng Cheng\textsuperscript{1}, 
Jonathan Ku\textsuperscript{1},
Qi Ding\textsuperscript{2},\\
Jiaqi Gu\textsuperscript{3},
Hanrui Wang\textsuperscript{2}, 
Hai "Helen" Li\textsuperscript{1},
Yiran Chen\textsuperscript{1}\\
\textsuperscript{1}Duke University, 
\textsuperscript{2}Massachusetts Institute of Technology,
\textsuperscript{3}Arizona State University\\
}

\begin{document}

\title{\name: \underline{Q}uantum Le\underline{g}alization and \underline{D}etailed \underline{P}lacement for Superconducting Quantum Computers}

\maketitle

\begin{abstract}
Noisy Intermediate-Scale Quantum (NISQ) computers are currently limited by their qubit numbers, which hampers progress towards 
fault-tolerant quantum computing. A major challenge in scaling these systems is crosstalk, which arises from unwanted interactions among neighboring components such as qubits and resonators. An innovative placement strategy tailored for superconducting quantum computers can systematically address crosstalk within the constraints of limited substrate areas.

Legalization is a crucial stage in placement process, refining post-global-placement configurations to satisfy design constraints and enhance layout quality. However, existing legalizers are not supported to legalize quantum placements. We aim to address this gap with \name, developed to meticulously legalize quantum components by adhering to quantum spatial constraints and reducing resonator crossing to alleviate various crosstalk effects.

Our results indicate that \name effectively legalizes and fine-tunes the layout, addressing the  quantum-specific spatial constraints inherent in various device topologies. By evaluating diverse NISQ benchmarks. \name consistently outperforms state-of-the-art legalization engines, delivering substantial improvements in fidelity and reducing spatial violation, with average gains of 34.4\x and 16.9\x, respectively.

\end{abstract}

\begin{IEEEkeywords}
Quantum computing, Placement, Legalization, Quantum design automation
\end{IEEEkeywords}

\section{Introduction}\label{sec:intro}
The rapid scaling of superconducting quantum computers (QCs) brings formidable challenges, notably managing crosstalk due to unintended electromagnetic interactions among quantum chip components \cite{SC, google, crosstalk}. Such interactions can significantly degrade computational fidelity by affecting gate operations when components with resonant frequencies are closely positioned \cite{crosstalk, para_g, substrate_limit, res_crosstalk, physical_process_2, device_crosstalk}. Furthermore, larger substrate sizes in superconducting qubits intensify electromagnetic coupling, leading to spurious modes that shorten coherence times and worsen crosstalk \cite{spurious, quantum_progress, dist_modeling}. A novel placement strategy, likening quantum device components to charged particles, offers a promising solution by achieving effective spatial and frequency isolation, while also addressing the challenges of substrate and inter-component crosstalk, thereby enhancing the scalability and fidelity of QCs \cite{qplacer}.

\begin{figure}[t]
  \centering
  \includegraphics[width=\linewidth]{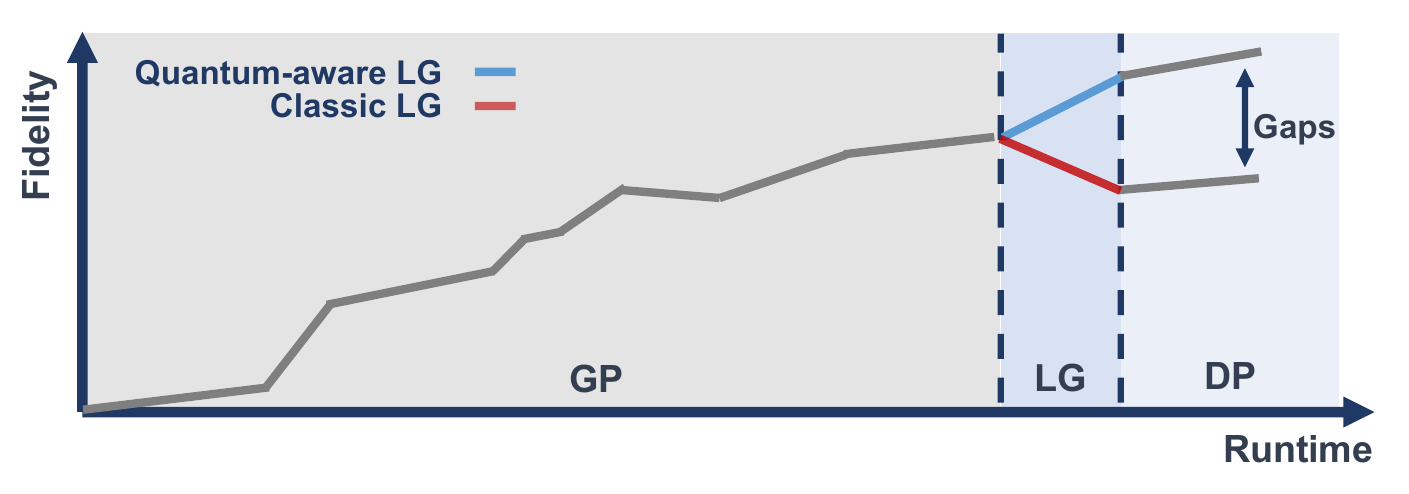}
  \vspace{-20pt}
  \caption{
  Impact of placement optimization stages on layout quality. Placement stages in sequence is global placement (GP, gray), legalization (LG, blue), and detailed placement (DP, light blue). The blue and red lines underscore the critical role of legalization. Despite its brief runtime, legalization considerably affects layout quality. Improper legalization can undermine the outcomes from GP, and these issues are often irreparable during DP. 
  }
  \label{fig:teaser}
  \vspace{-20pt}
\end{figure}

Despite advancements in quantum system placement, existing method primarily concentrates on the global placement (GP) stage, which merely determines rough locations for components. However, as quantum systems grow in complexity and scale, the subsequent stages of legalization (LG) and detailed placement (DP) become increasingly crucial. 
The objective of these stages is to resolve design rule violations and incrementally enhance the overall solution quality \cite{dreamplace}. Additionally, each component should be as close as possible to its original position determined during the GP to preserving the GP quality. 

\Cref{fig:teaser} demonstrates the relationship between layout quality versus placement optimization stages. The LG stage, despite its brief runtime, considerably affects layout quality. Traditional legalizers, illustrated by the red line and designed for classical systems, are inadequate for quantum placements as they fail to address quantum-specific challenges like crosstalk, focusing instead on eliminating overlaps and boundary issues. 
Such an improper legalization can undermine GP outcomes, and these deficiencies are typically irreparable during DP. 
Moreover, the partitioning of resonators is a promising technique to enhance the flexibility of the placement design \cite{qplacer}. However, the challenge is reintegrating these wire blocks without causing excessive crosses. Scattered wire blocks lead to numerous crossovers in resonators, necessitating the use of many airbridges \cite{airbridges}. This is problematic as airbridges reduce the fidelity of resonators \cite{cpw_cross}. Therefore, quantum-aware legalizers, as represented by the blue line in \Cref{fig:teaser}, are needed to resolve quantum spatial constraints and resonator crossings.

To address above problems and further improve the fidelity of quantum layout without sacrifice of area utilization, we present \name{}, a legalization and detail placement engine tailored for superconducting QCs. \name{} organizes legalization into two phases. Initially, it focuses on qubit legalization by ignoring the resonators, ensuring minimum spacing between qubits and minimal displacement to maintain the quality of GP.
Following qubit fixation, \name{} transitions to resonator legalization, focusing on the aforementioned integrity problem.
It ensures that wire blocks of each resonator are proximate to at least one other block of the same resonator, again with minimal displacement.
Subsequently, \name{} applies a window zoom to areas where theoretical crosstalk exists or resonator crossings occur, extracting and re-placing resonators to reduce crosstalk and resolve crossing points. All above processes are centralized around meeting specific quantum spatial constraints.

\noindent The contributions of this work are summarized as follows:
\setlist{nolistsep}
\begin{itemize}[leftmargin=*]
    \item To our knowledge, this research is the first to comprehensively address the challenges of legalization and detailed placement in quantum layout design, significantly enhancing quantum system fidelity and scalability.
    \item We introduce \name, a framework designed to meticulously legalize the quantum layout with adhering quantum spatial constraints, minimizing the resonator crosses and fine-tuning layout details to mitigate various crosstalk.
    \item To achieve this, \name systematically organizes the legalization process into stages for qubits and resonators. followed by the deployment of a detailed placer that identifies and refines regions with spatial violations.
\end{itemize}

\section{Background}

\subsection{Transmon Qubits and Couplers}\label{sec:qubit}

\begin{figure}[t]
  \centering
  \includegraphics[width=\linewidth]{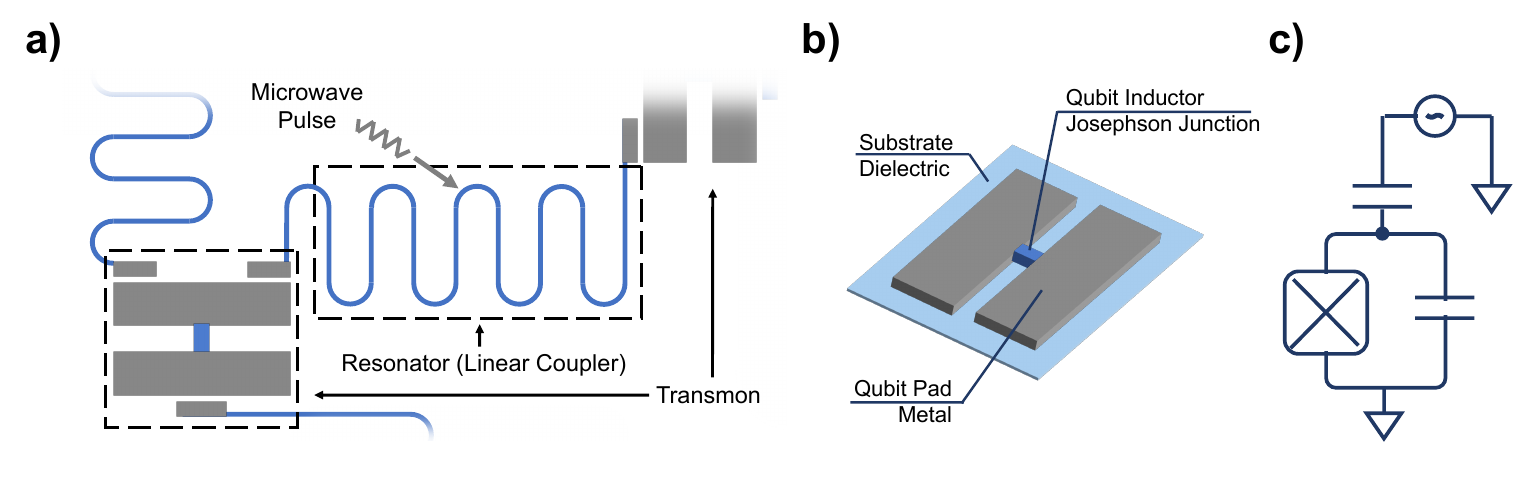}
  \vspace{-20pt}
  \caption{
  a) Transmon qubits coupled by resonators; two-qubit gates activated by applying/removing an off-resonant pulse to the resonator.
  \; b) Physical layout of a transmon qubit. 
  \; c) Circuit diagram of a fixed-frequency transmon qubit featuring a capacitor, Josephson junction, and microwave control line. 
  }
  \label{fig:bg_cQED}
  \vspace{-15pt}
\end{figure}

Transmon qubits are leading superconducting QC architectures \cite{quantum_progress, google, majority, CSAR}. These qubits are predominantly coupled using physical mechanisms such as capacitors, resonators (linear coupler), and tunable couplers \cite{cQED, intro_cQED, tunable_coupler}, with this work focusing on fixed-frequency transmons coupled by resonators \cite{SC, SC_1}, as in \Cref{fig:bg_cQED}-a. Each resonator functions as a quantum harmonic oscillator composed of a linear inductor and capacitor. The structure of a transmon qubit includes two metallic pads connected by a non-linear inductor (Josephson junction), forming a quantum anharmonic oscillator designed to emulate an atom-like energy spectrum with primary states: the ground state $|0\rangle$ and the first excited state $|1\rangle$ .

In these systems, single qubit gates are executed by modulating microwave voltage signals connected via a capacitor, detailed in \Cref{fig:bg_cQED}-c \cite{cQED}. 
For two-qubit gates, this architecture utilizes all-microwave-based methods that trigger qubit interactions through off-resonant pulses. These methods improve gate lifetimes, streamline control, and minimize crosstalk \cite{detune_flux}. \Cref{fig:bg_cQED}-a illustrates the control mechanism by applying an off-resonant pulse to the resonator, inducing phase shift in the qubits, thereby enabling gate operation\cite{rip_gate}.

\subsection{Challenges in Resonator Cross Points} \label{sec:cross}
Superconducting qubits require complex wiring setups to connect resonators, control lines, and measurement devices, facilitating couplings between qubits. As qubit arrays expand, maintaining accessible connection lines for internally located qubits becomes increasingly challenging, known as the "wiring problem." This issue is exacerbated when wire blocks cannot be effectively recombined post-partitioning, resulting in numerous crossing points \cite{qplacer, cQED}. Unlike traditional silicon integrated circuits, establishing compact wiring configurations in superconducting QCs is more complex due to the considerable decoherence from conventional multi-layer wiring methods used in silicon-based circuits \cite{multi_layer_problem}.

\begin{figure}[t]
  \centering
    \includegraphics[width=0.9\linewidth]{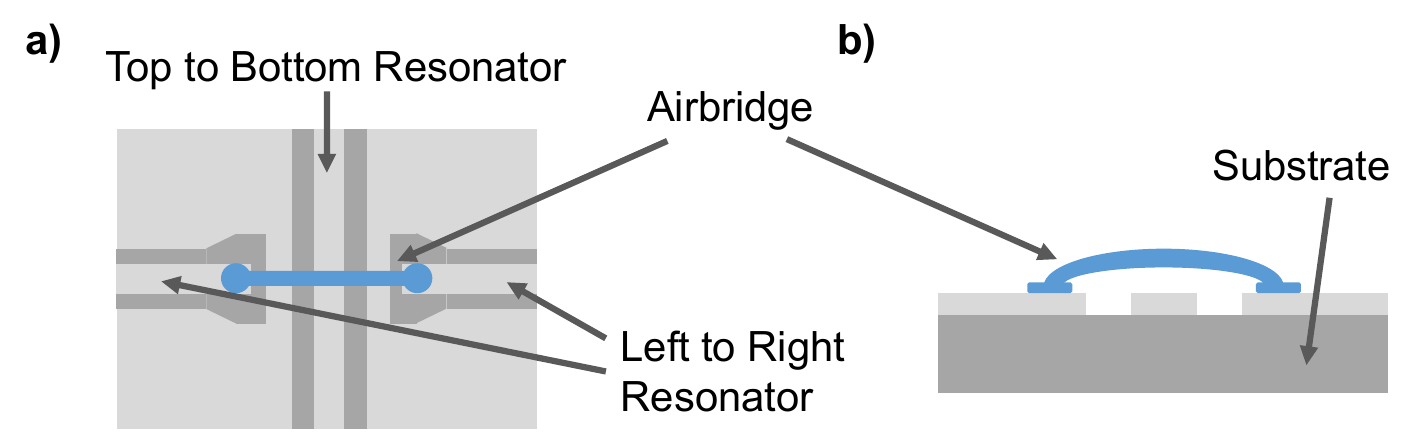}
  \vspace{-5pt}
  \caption{
  Airbridge diagram; A airbridge (blue) connects the signal lines of horizontal resonator from left to right, bridging over the vertical resonator.\; a) Top-view; \; b) Side-view, 
  }
  \label{fig:bg_cross}
  \vspace{-15pt}
\end{figure}

One practical solution for managing cross wiring in superconducting QCs is the use of airbridges. \Cref{fig:bg_cross} illustrates both top-view and side-view of airbridges design. Airbridges are monolithic microstructures designed to provide low-loss electrical connections over qubits and can be manufactured using established techniques \cite{multi_layer_cpw}. 
However, even with their benefits, airbridges should also be used with limits to address wire crossing issues because they can induce crosstalk, especially if intersecting resonators are not sufficiently detuned \cite{cpw_cross}. This underscores the importance of meticulous layout planning in superconducting QCs to minimize resonator crossing.

\section{\name Framework}
\subsection{Overview}
\name framework initiates with the qubit legalization, which strategically positions qubits to minimize displacement and maintain spatial compliance. Following this, Legalizer fixes the qubits and turns its target to resonators, aiming to optimize their integrity and reduce airbridge usage. After legalization, a detailed placement engine is activated, focusing on region with non-unified resonators and frequency hotspots. This section outlines the framework's objectives and details each step.

\subsection{Problem Formulation}\label{sec:problem}
The objective of \name is to legally position all quantum components and subsequently fine-tune their locations to minimize crosstalk impacts while preserving the GP quality. The mathematical formulation is presented as follows:

\noindent\textbf{Definition of Quantum Netlist:} A quantum netlist is defined as an undirected graph $G(Q, E)$, where each vertex $q$ corresponds to a qubit and each edge $e$ symbolizes a resonator coupling two qubits. Each edge $e_{ij}$ can be defined as tuples $(q_i, q_j, S_{ij})$, where $S_{ij}$ is the set of resonator wire blocks. Wire blocks within an edge are grouped into clusters if they physically touch, indicating integration and minimizing crossing points. A non-unified edge consists of multiple clusters and is represented as $(q_i, q_j, \{C_{ij}^1, \ldots C_{ij}^n\})$, where $C_{ij}^1\cup C_{ij}^2 \cup \ldots C_{ij}^n = S_{ij}$.

\noindent\textbf{Layout Constraints:}
\begin{itemize}[leftmargin=*]
    \item \textit{Non-overlapping:} To prevent overlap, the position of each quantum component $i$ and $j$, where $i, j \in G$, must satisfy:
    
    \begin{equation}
        |x_{i} - x_{j}| \geq \frac{w_{i} + w_{j}}{2}, \quad |y_{i} - y_{j}| \geq \frac{h_{i} + h_{j}}{2}
    \end{equation}
    
    where $(w, h)$ represent the dimensions of each component's bounding polygon.
    \item \textit{Border Constraints:}  Each component must remain within the defined borders $(W, H)$:
    
    \begin{equation}
    \frac{w_{i}}{2} \leq x_{i} \leq W - \frac{w_{i}}{2}, \quad \frac{h_{i}}{2} \leq y_{i} \leq H - \frac{h_{i}}{2}
    \end{equation}
    
\end{itemize}

\noindent\textbf{Objectives:}
\begin{itemize}[leftmargin=*]
\item \textit{Minimize Cluster Count:} Aim to reduce the total number of clusters across all edges to enhance layout quality. The ideal scenario for each edge is a single cluster, $|C_e| = 1$, which indicates unified resonator:

\begin{equation}
\text{Minimize} (\sum_{e \in E} |C_e|)
\end{equation}

\item \textit{Minimize Frequency Hotspot Proportion ($P_h$):} This metric quantifies potential crosstalk risks, identifying areas where component frequencies are closely matched and spatially proximate, thus requiring mitigation \cite{qplacer}:

\begin{equation}
    \text{Minimize } (P_{h} = \frac{\sum_{i, j \in G} (p_i \cap p_j) \cdot d_c(p_i, p_j) \cdot \tau(\omega_i, \omega_j, \Delta_c)}{\sum_{n \in G} w_n*h_n})
\end{equation}

Here, $p_n$ represents the polygon of components $n$, $p_i \cap p_j$ is the intersection length between two polygons, $d_c(p_i, p_j)$ is the centroid distance, and $\tau$ is a function assessing frequency proximity according to defined component frequency $\omega$ and predefined threshold $\Delta_c$.
\end{itemize}

\noindent\textbf{Input:} A quantum netlist $G$ with initial positions $(x, y)$ for each component from the global placement.

\noindent\textbf{Output:} Optimized positions $(\hat{x}, \hat{y})$ that adhere to the stated objectives and constraints.

\subsection{Qubit Legalization} \label{sec:q_lg}
\begin{figure}[t]
  \centering
  \includegraphics[width=0.85\linewidth]{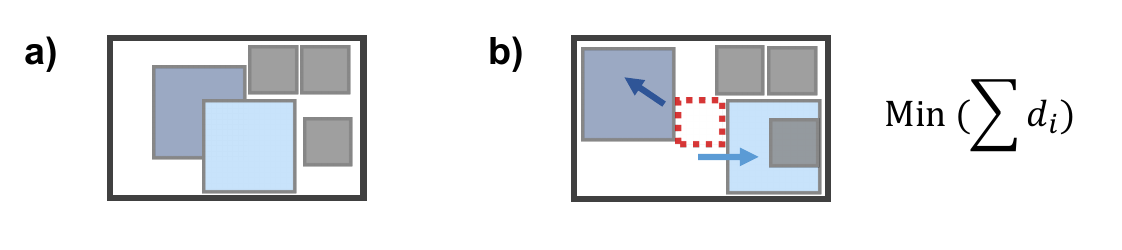}
  \vspace{-15pt}
  \caption{
  Qubit legalization, black box represent the layout border, qubits (blue) and resonator segments (gray) are color-coded by frequency; \;
  \textbf{a):} GP positions; \;
  \textbf{b):} Post-qubit legalization (red dot box depicts minimum spacing constraint, arrows show the displacement)
    }
  \label{fig:method_q_lg}
  \vspace{-20pt}
\end{figure}
Qubit legalization is first carried out, temporarily disregarding resonators at this phase. \name defines the size of the resonator segments as the standard cell. Consequently, qubits can be analogue to Macros in VLSI design, as their size significantly exceeds that of the segments (standard cell).
We adopt a macro legalization strategy using linear programming \cite{macro}. This method constructs horizontal and vertical constraint graphs with macros (qubits) as nodes and permissible movements as arcs, utilizing dual min-cost flow algorithms to minimize qubit displacement. The objective is to minimize total displacement:

\begin{equation}
    \text{Minimize} (\sum_{i\in Q} d_i)
\end{equation}
where $d_i$ denotes the displacement of qubit $i$ from its initial positions, striving to maintain qubits as close to their GP locations as possible to preserve the initial logical layout.

Padding technique in the GP stage helps meet quantum spatial requirements but involve trade-offs: larger padding reduces area utilization, whereas less padding increases the risk of qubit crosstalk \cite{qplacer, para_g}. 
To optimize padding without sacrificing fidelity, we shift part of the spacing task to the qubit legalization phase. Given that resonators operate at higher frequencies than qubits, they effectively isolate and mitigate inter-qubit crosstalk \cite{cQED}.
Thus, it is crucial to maintain at least one standard cell size spacing between adjacent qubits during legalization. 
This minimum spacing is enforced as a constraint in our solver, and uses a greedy method to dynamically adjust spacing. 
The solver starts with stringent constraints and relaxing them only when necessary to achieve a compact yet compliant layout. This iterative adjustment is crucial for densely packed qubit arrays. 
\Cref{fig:method_q_lg}-a displays the component positions from the GP phase, while \Cref{fig:method_q_lg}-b illustrates the layout post-qubit legalization, emphasizing the enforced separation between qubits (red dot box) and minimized displacement (arrows).

\subsection{Resonator Legalization} \label{sec:res_lg} 
\begin{figure}[t]   
  \centering
  \includegraphics[width=0.8\linewidth]{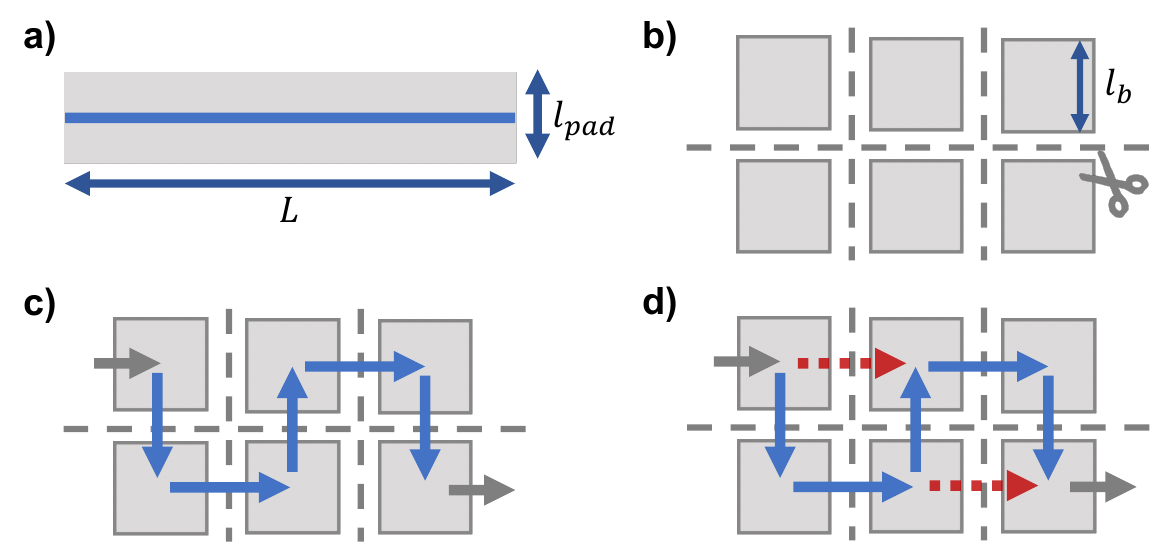}
  \vspace{-5pt}
  \caption{
  pseudo connection in defining netlist
  \textbf{a):} Padded resonator with wirelength $L$ and padding length $l_{pad}$;
  \;
  \textbf{b):} Reshaped resonator from \textbf{(a)} into a compact rectangle, partitioned into segments of size $l_b$, retaining frequency consistency as indicated by color. Here $n=6$;
  \;
  \textbf{c):} Net connection of wire blocks (blue arrow indicates the connection between blocks, gray arrow indicates the connection between blocks and qubits);
  \;
  \textbf{d):} Enhanced net connection with pseudo connects (red dot arrow is the enhanced connections to lead a legalization friendly GP layout)
  }
  \label{fig:method_pseudo}
  \vspace{-15pt}
\end{figure}
\noindent\textbf{Pseudo Connection:\xspace} Resonator partitioning is a critical technique used during GP phase of superconducting QCs to enhance flexibility and scalability by addressing resonator area overhead \cite{qplacer}. This process involves initially reshaping resonators into compact rectangles of equivalent area. These are then segmented based on a predefined block size \(l_b\), as illustrated in \Cref{fig:method_pseudo}-a and b. The area relationship between the padded resonator and segments is described by:

\begin{equation}\label{eq:res_relation}
    l_{pad} \cdot L =  n \cdot l_b^2
\end{equation}
where \(L\) is the wire length of the resonator, \(n\) is the number of wire blocks post-partition. This strategy preserves the resonator's fundamental frequency properties while allowing for individual placement of segments within substrate constraints. Note: the purpose of these blocks is solely to reserve layout space for resonators; the detailed routing within the reserved space is beyond the scope of this study.

In \cite{qplacer}, wire blocks are connected in a snake-like pattern, resulting in an elongated line configuration rather than the desired rectangular shape (\Cref{fig:method_pseudo}-c). This linear arrangement originates from the density objectives in the GP, which can spread out cells if not constrained by network connections. This layout results in two critical issues:
\textbf{1) Legalization Challenges:} Linear arrangement of wire blocks causes significant overlap and restricted space, leading to substantial displacement during legalization and disruption of the GP positions.
\textbf{2) Crosstalk Risk:} A linear pattern increases the perimeter of the reserved space, elevating the potential for crosstalk.

To mitigate these issues, this study introduces a "pseudo connection" strategy, where wire blocks are interconnected with all neighboring segments during netlist creation. This approach, depicted by the red arrows in \Cref{fig:method_pseudo}-d, fosters a more rectangular resonator layout post-GP. This configuration greatly simplifies the challenges of resonator legalization and maintains a compact, design-coherent placement.

\begin{algorithm}[b]
\caption{Resonator Integration-aware Legalization} \label{alg:res_lg}
\begin{algorithmic}[1]
\scriptsize

\Require Qubit legalization solution $\text{p}(i)$ for all movable quantum components $I$, $\forall i \in I$; 
\quad Area of substrate $(X, Y)$;
\quad Segment list $S_e$ and cluster list $C_e$ for each resonator $e, \forall e \in E$, and 

\noindent $C_{e}^1\cup C_{e}^2\ldots C_{e}^n = S_e$; 
\quad Adjacent available bin update function $f(\cdot)$; \quad Displacement calculator $d(\cdot)$

\Ensure Legalize $E$ and Minimize the number of clusters $(|C_e|)$, $\forall e\in E$ (Integration-aware)

\State $B \gets (X, Y) \quad\quad\quad\quad\quad\quad\quad\quad\quad\quad 
\triangleright$ Obtain all the Bins
\State $B_{f} \gets \text{p}(q), \forall q \in Q \; \text{and}\;(X, Y)
\quad \triangleright$ Obtain fixed Bins
\State $B_{a} \gets  B - B_{f} \quad\quad\quad\quad\quad\quad\quad\quad\quad
\triangleright$ Obtain available Bins
\For{$e \in E$}
    \State $B_{aa} \gets \emptyset \quad\triangleright$ Initialize adjacent available Bins
    \For{$s \in S_e$}
        \If{$B_{aa} == \emptyset$}
            \State $\text{p}(\hat{b}) \gets \text{minimum}(d(\text{p}(s), \text{p}(b)), b \in B_a $
        \Else
            \State $\text{p}(\hat{b}) \gets \text{minimum}(d(\text{p}(s), \text{p}(b)), b \in B_{aa}$
        \EndIf
        \State $\text{p}(s) \gets \text{p}(\hat{b})
        \quad\quad\quad\quad\quad \triangleright$ Legalize segment
        \State $B_a \gets B_a-\text{p}(s)
        \quad\quad\quad\quad \triangleright$ Update $B_{a}$
        \State $B_{aa}\gets f(B_{aa}, B_a, \text{p}(s)) \quad \triangleright$ Update $B_{aa}$
    \EndFor
\EndFor

\end{algorithmic}
\end{algorithm}

\begin{figure}[t]
  \centering
  \includegraphics[width=0.9\linewidth]{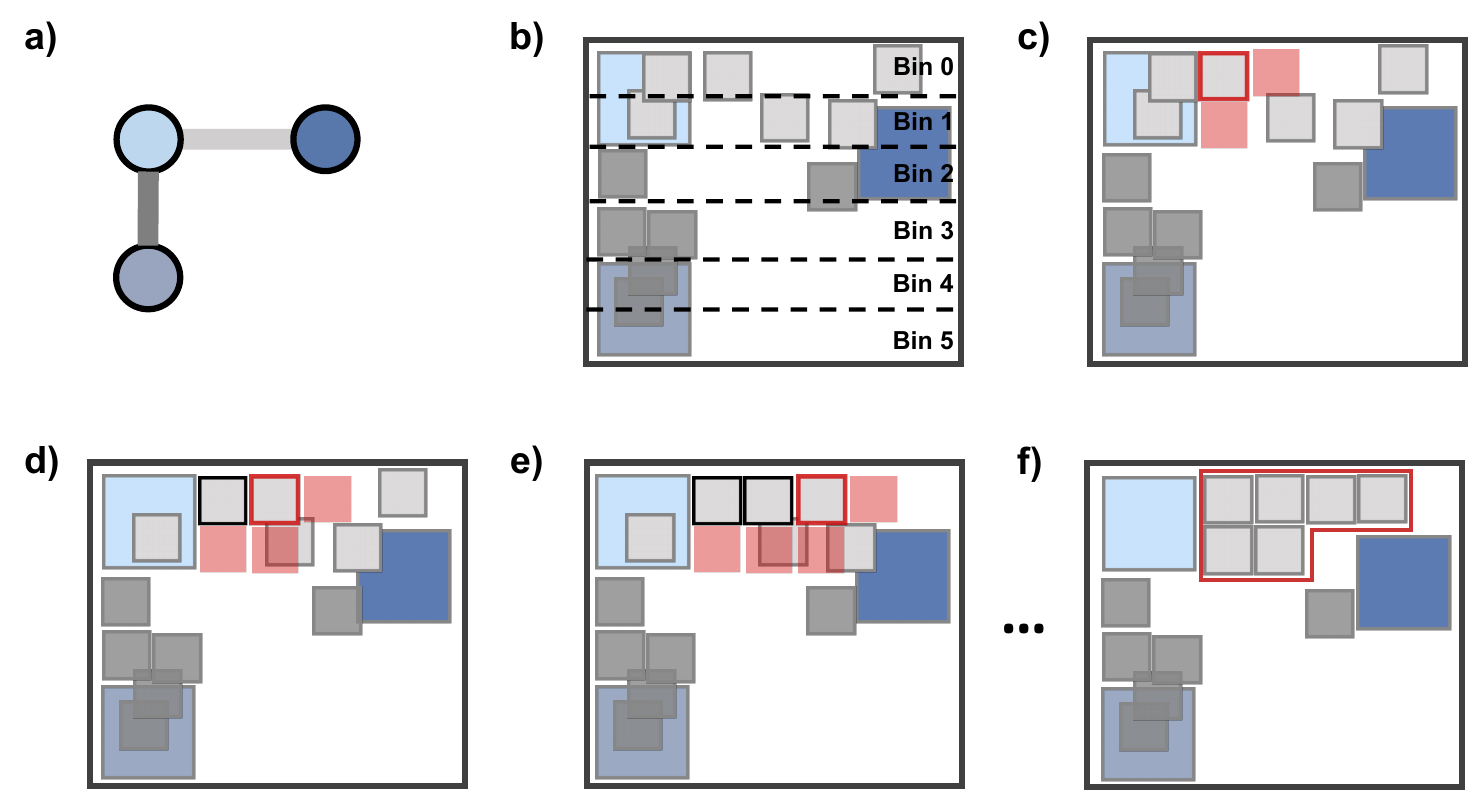}
  \vspace{-5pt}
  \caption{
  Resonator Integration-aware legalization, components are color-coded by frequency;
  \;
  a) Connectivity topology;
  \;
  b) Given qubit legalization positions (bin-aided cell search);
  \;
  c): The first wire block (red box) from the resonator (gray) is legalized with minimum displacement. Adjacent available spaces to the legalized blocks from this resonator (gray) is highlighted (light red);
  \;
  d): Legalizing the next wire block (red box) from the same resonator and the adjacent available spaces (light red) is updated;
  \;
  e): Legalization of this resonator continues (red box) and adjacent available spaces keep (light red) updating.
  \;
  f): All the wire blocks from this resonator (light gray) are legalized (red box), move to next resonator (dark gray)
  }
  \label{fig:method_res_lg}
  \vspace{-20pt}
\end{figure}

\noindent\textbf{Integration-aware legalization:\xspace} After positioning the qubits, the focus shifts to the legalization of segmented resonator blocks using a modified Tetris-like methodology \cite{tetris}. The primary objective is to minimize cluster count, as outlined in \Cref{sec:problem}. A significant challenge in this phase is efficiently handling the collection of all legalized cells and available spaces, especially at scale. 
To improve scalability and runtime efficiently, we adopted a bin-aided indexing approach \cite{bin-aided}, which organizes cells into hierarchical bins along the y-axis rather than flattened arrays, reducing cell query operations to \(O(\log n)\). This strategy significantly narrows the search region and reduces the overhead associated with placing legalized cells into specific bins, as depicted by dashed lines in \Cref{fig:method_res_lg}-b. In subsequent subfigures, auxiliary lines are omitted for clarity.

\Cref{fig:method_res_lg} visualizes an example of resonator legalization process (light gray one). 
The legalization begins in \Cref{fig:method_res_lg}-c, where the first wire block is legalized in an optimal position to minimize displacement, marked by a red box. Adjacent potential locations for the next wire block are highlighted in light red. 
In \Cref{fig:method_res_lg}-d, previously placed segments are denoted by black boxes, and the block currently being legalized is in a red box, positioned in the most favorable adjacent space. The selection of this space is determined by the least displacement. And adjacent available space is keeping updated. 
This iterative process continues until all segments of this resonator are legalized, as depicted in \Cref{fig:method_res_lg}-e and f.

\Cref{alg:res_lg} outlines the details of resonator legalization process using a bin-aided design. Initially, available bins (\(B_a\)) and adjacent available bins (\(B_{aa}\)) are identified. The process prioritizes placing wire blocks into \(B_{aa}\) spaces with the smallest displacement. If \(B_{aa}\) is empty, blocks are then placed into the nearest available space in \(B_a\), also with the smallest displacement. This sequence continues until all segments from a resonator are legalized before moving to the next resonator. This integration-aware methodology enhances layout performance by effectively managing segment interactions within the layout minimizing the usage of airbridges.

\begin{figure}[t]
  \centering
  \includegraphics[width=\linewidth]{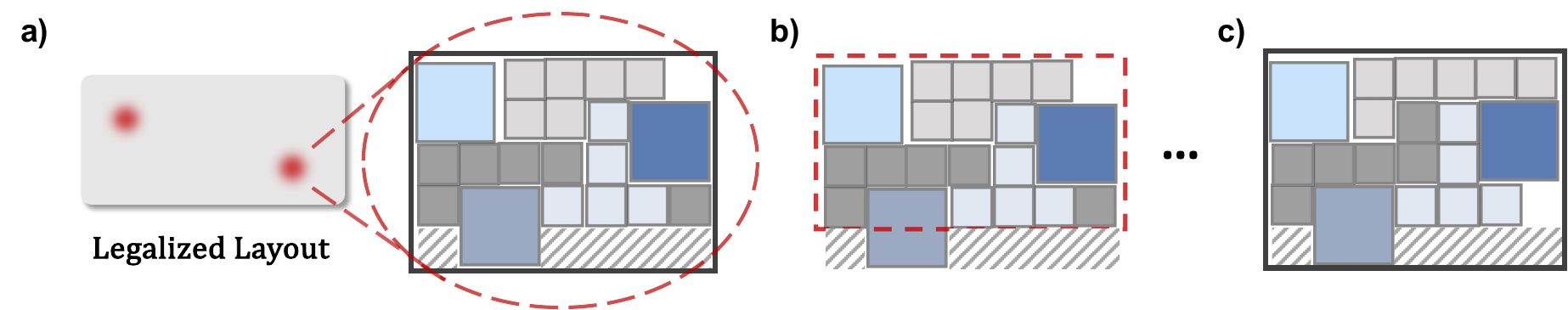}
  \vspace{-15pt}
  \caption{
    Detailed placement
    \;
    \textbf{a):} Identify regions with constraint violations, noted by red dots and zoomed in on the right side (areas with slashed lines are unavailable).
    \;
    \textbf{b):} Construct a focused window (outlined by a red dashed box).
    \;
    \textbf{c):} Extract and reposition resonators to resolve spatial constraint violations.
    }
  \label{fig:method_dp}
  \vspace{-20pt}
\end{figure}

\begin{algorithm}[b]
\caption{Detailed Placement} \label{alg:dp}
\begin{algorithmic}[1]
\scriptsize

\Require Legalization solution $\text{pos}(i)$ for all movable quantum components $I$, $\forall i \in I$; 
\; Segment list $S_e$, cluster list $C_e$, frequency hotspots $H_e$ for each resonator $e, \forall e \in E$, and $C_{e}^1\cup C_{e}^2\ldots C_{e}^n = S_e$; 
\; Maze routing $M(\cdot)$

\Ensure Optimize $E$ by minimizing the number of clusters $\sum (|C_e|)$ and frequency hotspots $\sum (H_e)$ in the window $W$, $\forall e\in W$

\State $E_c \gets |C_{e}|>1, \forall e\in E \quad \triangleright$ Obtain non-unified resonators
\State $E_h \gets H_{e}>0, \forall e\in E \quad\; \triangleright$ Obtain resonators with hotspots
\For{$e \in E_c \cup E_{h}$}
    \State $E_e \gets e, (x_i, y_i) \quad\quad \triangleright$ Obtain adjacent resonators 
    \State $W_{e} \gets \{e, E_e\} \quad\quad\; \triangleright$ Construct window 
    \State $\hat{e} \gets M(W_{e}) \quad\quad\quad \triangleright$ Optimize window.
    \If{$|C_{e}| < |C_{\hat{e}}|\; \text{and} \; H_e < H_{\hat{e}}$}
        \State $\text{pos}(i) \gets \hat{e}, \forall i\in W_e \quad\triangleright$ update positions
    \EndIf
\EndFor

\end{algorithmic}
\end{algorithm}

\subsection{Detailed Placement} \label{sec:dp}
The detailed placement engine in this study focuses on optimizing resonator positions without altering the positions of qubits. \Cref{fig:method_dp} outlines the detailed placement (DP) procedure. Initially, the engine scans the entire legalized layout to identify resonators with multiple clusters ($|C_e| > 1$) and frequency hotspots $H_e$, marked by red dots in \Cref{fig:method_dp}-a left.

Subsequently, the detail placer addresses these issues one by one. For example, \Cref{fig:method_dp}-a right shows a zoomed-in view of a non-unified resonator. A processing window $W$ (red dotted box in \Cref{fig:method_dp}-b) is defined around the problematic resonator and its adjacent resonators to facilitate focused repositioning. This window includes the minimum bounding box necessary for the proximity of these resonators to potentially necessitate their repositioning. In \Cref{fig:method_dp}-b, adjacent resonators are displayed in light gray and light blue, and their wire blocks are extracted for rerouting. Maze routing establishes efficient paths for these resonators, optimizing connectivity and avoiding blocked cells.

After optimization, the window region $W$ is thoroughly reviewed to verify the fidelity of each resonator. If the cumulative cluster count $\sum|C_e|$ or frequency hotspots $\sum H_e$ post-optimization exceeds those from the legalization phase, the placements from the detailed stage are discarded. This ensures that the detailed placement improves the layout configuration by strictly adhering to constraints. Details are in \Cref{alg:dp}.

\section{Evaluation Methodology}
\noindent\textbf{Implementation:}
\name was developed using Python, utilizing PyTorch for optimizers and APIs, and C++ for low-level operations, building upon the open-source placer \cite{dreamplace}. Experiments were conducted on a Linux machine with an Intel(R) Xeon(R) CPU E5-2687W v4 @3.00GHz.
The architecture of \name comprises two main components: \name-LG, the proposed quantum legalizer addressing both qubit (\Cref{sec:q_lg}) and resonator legalization (\Cref{sec:res_lg}), and \name-DP, the proposed detailed placer (\Cref{sec:dp}). For details on qubit geometry features, please refer setups in \cite{qplacer}.

\noindent\textbf{Benchmarks:}
The evaluation conducts using a variety of quantum device connectivity topologies and NISQ benchmarks, detailed in \Cref{tab:benchmark}. These topologies, ranging from 25 to 127 qubits, include designs used in industrial applications and those optimized for algorithmic efficiency.

\noindent\textbf{Baselines:}
We conduct comparative evaluations of baselines to assess the performance of our legalizer and detailed placer. For consistency, all comparisons are based on the same GP positions with pseudo connections, as detailed in \Cref{sec:res_lg}.
\begin{itemize}[leftmargin=*]
    \item \textbf{Tetris} Utilizes macro legalizer \cite{macro} for qubits and Tetris legalizer for resonators \cite{tetris}.
    
    \item \textbf{Abacus} Employ macro legalizer \cite{macro} for qubits, paired with an Abacus legalizer \cite{abacus} for resonators.
    
    \item \textbf{Q-Tetris} Replaces the macro legalizer with our qubit legalization approach from \Cref{sec:q_lg}, while maintaining Tetris approach for resonators.
    
    \item \textbf{Q-Abacus} Substitutes the macro legalizer with our qubit legalization method from \Cref{sec:q_lg} and uses Abacus method for resonators.

\end{itemize}

\begin{table}[t] 
  \centering
  \scriptsize
  \caption{TOPOLOGIES AND BENCHMARKS}
    \vspace{-5pt}
  \begin{tabular}{lcl}
    \toprule
    Topology & Qubits & Description \\
    \midrule
    Grid      & 25 & Quantum error correction friendly architecture\cite{google, surface_code}\\
    Heavy Hex & 27  & Falcon processor from IBM \cite{ibmq}\\
    Heavy Hex & 127 & Eagle processor from IBM \cite{ibmq}\\
    Octagon   & 40  & Aspen-11 processor from Rigetti \cite{rigetti}\\
    Octagon   & 80  & Aspen-M processor from Rigetti \cite{rigetti}\\ 
    Xtree     & 53  & Pauli-String efficient architecture in Level 3 \cite{xtree}\\

    \midrule
    \midrule
    
    Benchmark & Qubits & Description \\
    \midrule
    BV    &4, 9, 16& Bernstein-Varzirani (BV) algorithm \cite{bv}\\
    QAOA  &4 & Quantum Approximate Optimization Algorithm \cite{QAOA} \\
    Ising &4 & Linear Ising model simulation of spin chain \cite{ising}\\
    QGAN  &4, 9 & Quantum Generative Adversarial Network \cite{QGAN}\\

  \bottomrule
  
\end{tabular}
\label{tab:benchmark}
\vspace{-15pt}
\end{table}

\begin{figure*}[t!]
    \centering
    \includegraphics[width=0.95\linewidth]{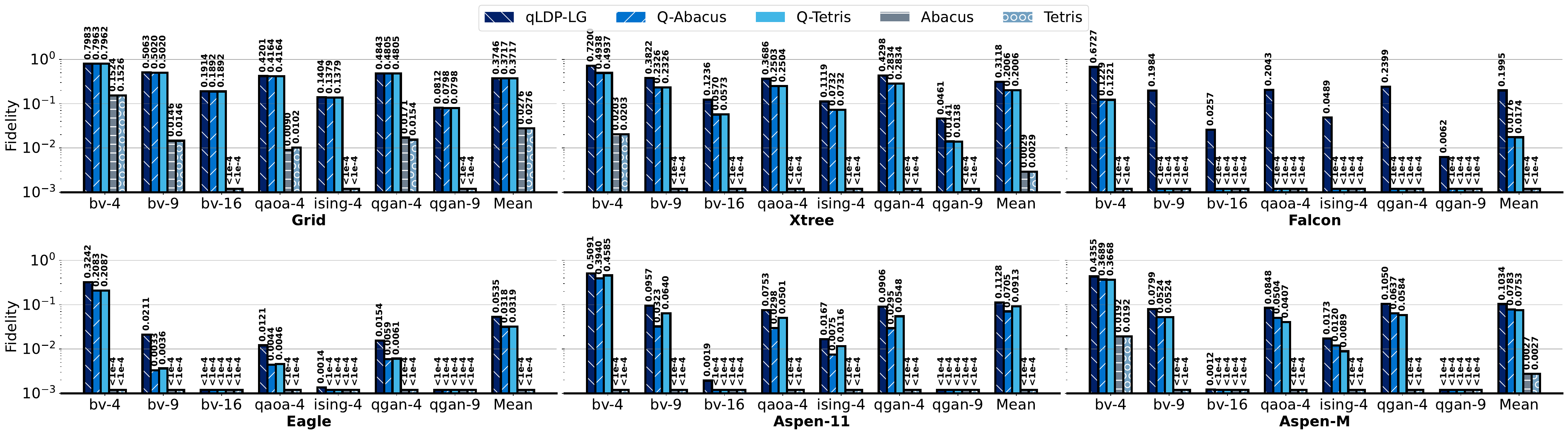}
    \vspace{-5pt}
    \caption{Fidelity estimation from various legalization strategies. A higher fidelity value indicates a better performance. Across all benchmarks, \name{} consistently outperforms the baselines, demonstrating its superior efficacy in maintaining higher fidelity levels in layout results.}
    \label{fig:exp_fidelity}
    \vspace{-15pt}
\end{figure*}

\noindent\textbf{Metrics:} 
To assess crosstalk susceptibility in our experiments, we analyse the layout quality from two perspectives: Program fidelity and Frequency hotspots proportion. 

\noindent\textbf{(1) General algorithm program fidelity:} Program fidelity $\mathcal{F}$ is estimated using three components to assess the worst-case fidelity of a benchmark affected by crosstalk and decoherence noises, following a similar approach to \cite{crosstalk_ding, google}:

\begin{equation} \label{eq:fidelity}
    \mathcal{F} = 
    \Pi_{q \in Q} (1-\epsilon_q) \cdot 
    \Pi_{g \in G}(1-\epsilon_g) \cdot 
    \Pi_{e \in E}(1-\epsilon_{e})
\end{equation}
where $\epsilon_q$ accounts for qubit errors from single and two-qubit gates and decoherence (modeled using decay constants $T_1$ and $T_2$). 
Crosstalk qubit error $\epsilon_g$ represents errors from qubits with spatial constraint violations, akin to being linked by a direct capacitive coupling. 
This error results from Rabi oscillations, periodic energy exchanges between the qubits, driven by their effective coupling strength $g_{\text{eff}}$. The transition probability is modeled as $\Pr[t] = \sin^2(g_{\text{eff}}t)$, and the corresponding crosstalk error for idle qubits is \cite{gate_time1}:

\begin{equation}
    \epsilon_g(\Delta,t) = 1-\sin{(g_{\text{eff}}(\Delta)t)}^2. 
    \label{eq:eg}
\end{equation}

\noindent Similarly, $\epsilon_e$ accounts for crosstalk errors among resonators under spatial violations or crossing points, similar to $\epsilon_g$.
The parasitic capacitance at each crossing point is set at 3.5 fF, as simulated with AWR Microwave Office \cite{AWR}. The parasitic capacitance for spatial violation is depends on adjacent length.
\textit{Note: these fidelity calculations apply only to actively engaged physical qubits (mapped) and resonators in the layout, as errors in inactive elements do not affect overall program fidelity.}

\noindent\textbf{(2) Frequency Hotspot Proportion:} As detailed in \Cref{sec:problem}, $P_h$ is used to quantifies crosstalk potential; An related metric is $H_Q$ which counts \#qubits under the crosstalk.

\section{Experiment Results}

\noindent\textbf{Legalization Fidelity:\xspace}
\Cref{fig:exp_fidelity} presents the worst-case overall fidelity for various legalization strategies, as quantified by the noise model in \Cref{eq:fidelity}. We evaluated each topological layout by performing 50 mappings of a benchmark program, with each bar in the figure representing the average fidelity.

Traditional legalizers like Abacus and Tetris, which overlook quantum spatial constraints such as minimum qubit spacing and cross minimization, show a marked decrease in fidelity, especially in complex topologies. Notably, Falcon exhibits lower fidelity than Xtree, which has nearly double the qubits. This outcome may stem from the Hex topology's fewer edges, which can result in deeper transpiled circuits.

To isolate the contributions of our qubit and resonator legalizers, we introduced Q-Tetris and Q-Abacus. These hybrid legalizers integrate our qubit legalizer with classical cell legalizers, achieving fidelity improvements of 22.9\x and 23.5\x, respectively, compared to their classical counterparts.

Enhancements are more pronounced with our resonator legalizer. \name-LG significantly outperforms traditional methods, improving fidelity by 1.5\x, 1.46\x, 34.4\x, and 34.4\x over Q-Abacus, Q-Tetris, Abacus, and Tetris, respectively. The success of \name-LG is attributed to its heuristic approach that optimally positions components to address quantum constraints.

\begin{figure}
    \centering
    \includegraphics[width=\linewidth]{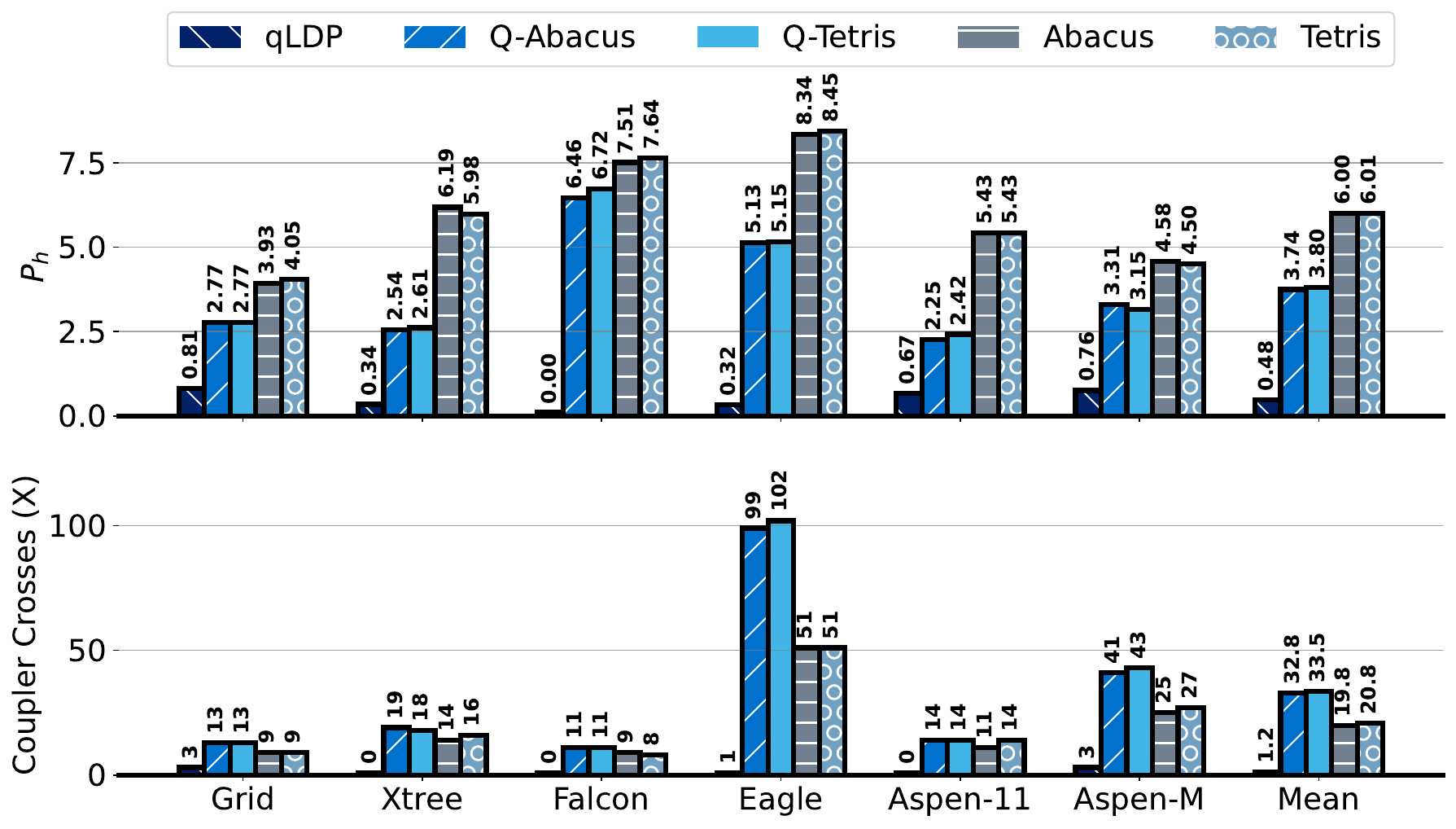}
    \vspace{-15pt}
    \caption{
    A comparison of different legalizers utilizes three metrics: Average program fidelity, frequency hotspots proportion $P_h$, and resonator crossings (\crossing), with lower values preferred for latter two.
    }
    \label{fig:exp_hotspots}
    \vspace{-10pt}
\end{figure}

\noindent\textbf{Crossing Points and Frequency Hotspots:\xspace}
\Cref{fig:exp_hotspots} presents the proportion of frequency hotspots ($P_h$) and the number of cross points (\crossing) under various legalization strategies. The data demonstrate that program fidelity is inversely related to $P_h$, validating its utility for assessing layout quality. Furthermore, it is hard to observe a strong correlation between the number of crosses and $P_h$, underscoring the non-localized nature of resonator crosstalk which impacts overall layout fidelity. While Q-abacus and Q-tetris effectively reduce $P_h$, they increase $X$ by 1.65\x and 1.61\x, respectively, over their classical counterparts.

\name{} significantly outperforms traditional legalization engines in achieving spatial isolation, maintaining an average violation rate of only 1.43\%. In contrast, classic legalizers exhibit higher average hotspot proportions of 3.74\%, 3.8\%, 6.0\%, and 6.01\%, making \name{} approximately 2.6\x to 4.2\x more effective at reducing crosstalk-related violations. Additionally, \name achieves significant improvements in minimizing resonator crossing ranging from 6.0$\sim$9.9\x compared to baselines.

\begin{table}[t!] 
  \centering
  \caption{Comparison of Legalization Time. \( t_q \) for qubits and \( t_e \) for resonators, measured in milliseconds (ms).}

  \scriptsize
  \vspace{-5pt}
  \setlength{\tabcolsep}{4pt} 
  \begin{tabular}{c|cc|cc|cc|cc|cc}
    \toprule
    \multirow{2}{*}{\tiny{Topology}} & \multicolumn{2}{c|}{\name-LG} & \multicolumn{2}{c}{Q-Abacus} & \multicolumn{2}{|c}{Q-Tetris} & \multicolumn{2}{|c}{Abacus} & \multicolumn{2}{|c}{Tetris} \\
     & \Q & \C & \Q & \C & \Q & \C & \Q & \C & \Q & \C \\
    \midrule
    Grid        & 1.62 & 1.11 & 1.41 & 0.85 & 1.35 & 0.74 & 0.62 & 0.80 & 0.88 & 0.70 \\
    Xtree       & 4.98 & 1.62 & 4.43 & 1.22 & 4.49 & 1.07 & 2.16 & 1.12 & 2.56 & 0.93 \\
    Falcon      & 2.27 & 0.87 & 1.54 & 0.53 & 1.53 & 0.42 & 0.68 & 0.48 & 0.98 & 0.39 \\
    Eagle       & 23.69 & 5.75 & 25.14 & 4.27 & 25.69 & 3.91 & 12.95 & 3.57 & 13.76 & 3.06 \\
\tiny{Aspen-11} & 3.33 & 1.54 & 3.02 & 1.11 & 2.99 & 0.97 & 1.49 & 0.98 & 1.86 & 0.87 \\
\tiny{Aspen-M}	& 10.81 & 3.71 & 10.51 & 2.56 & 10.47 & 2.32 & 5.44 & 2.23 & 6.19 & 2.01\\
    Mean        & 7.78 & 2.43 & 7.68 & 1.76 & 7.75 & 1.57 & 3.89 & 1.53 & 4.37 & 1.32\\
  \bottomrule
\end{tabular}
\label{tab:exe_time}
\vspace{-15pt}
\end{table} 

\noindent\textbf{Runtime:\xspace}
We evaluated the runtime of our proposed quantum legalizers, \name-LG, against established baselines. \Cref{tab:exe_time} presents the legalization times, measured in milliseconds, for qubits ($t_q$) and resonators ($t_e$). As detailed in \Cref{tab:exe_time}, \name-LG exhibits a balance between efficiency and effectiveness, with mean legalization times for qubits and resonators being within a competitive range of the fastest baseline methods. Notably, the $t_q$ for \name-LG, Q-Abacus, and Q-Tetris are generally longer than those for Abacus and Tetris, due to the aggressive initial minimum spacing settings that require iterative adjustments, as discussed in \Cref{sec:q_lg}. Overall, \Cref{tab:exe_time} demonstrates the efficacy of our algorithms in optimizing quantum circuit layouts, ensuring legal placements without sacrificing operational efficiency or design integrity.

\begin{table}[t!] 
  \centering
  \caption{Detailed Placement Evaluation, $I_\text{edge}$ is the number of unified resonators over total resonators (higher the better), X for resonator crossings, $P_h(\%)$ for the proportion of frequency hotspots, and $H_Q$, the number of qubits affected by hotspots, with lower values preferred for the last three metrics. }
  \scriptsize
  \vspace{-5pt}
  \setlength{\tabcolsep}{4pt} 
  \begin{tabular}{c|c|cccc|cccc}
    \toprule
    \multirow{2}{*}{Topology} & \multirow{2}{*}{\#Cells} & \multicolumn{4}{c}{\name-LG} & \multicolumn{4}{|c}{\name-DP} \\
     & & $I_\text{edge}$ & X & $P_h(\%)$ & $H_Q$ & $I_\text{edge}$ & X & $P_h(\%)$ & $H_Q$ \\
    \midrule
    Grid   &  490 & 37/40  & 3 & 1.38 & 11 & 37/40 & 3 & 0.81 & 5 \\
    Xtree  &  660 & 47/52  & 5 & 1.37 & 20 & 52/52 & 0 & 0.34 & 10 \\
    Falcon &  354 & 28/28  & 0 & 0.92 & 8  & 28/28 & 0 & 0 & 0\\
    Eagle  & 1801 & 142/144 & 2 & 1.27 & 68 & 143/144 & 1 & 0.32 & 15 \\
\tiny{Aspen-11} &  598 & 46/48  & 2 & 0.91 & 20 & 48/48 & 0 & 0.66 & 9 \\
\tiny{Aspen-M}	& 1310 & 98/106 & 8 & 2.71 & 50 & 103/106 & 3 & 0.76 & 14 \\
  \bottomrule
\end{tabular}
\label{tab:dp_eva}
\vspace{-10pt}
\end{table}

\noindent\textbf{Analysis of Detailed Placement:\xspace}
We analyze the performance of \name-DP in comparing with the layout of \name-LG in \Cref{tab:dp_eva}. 
\name-DP consistently outperforms \name-LG, achieving better or equivalent $I_{\text{edge}}$ scores across all configurations, which suggests a superior ability to maintain optimal connections within the circuits. Notably, DP not only meets but often surpasses LG in minimizing the number of cross and $P_h$, indicating a robust adherence to design constraints. 
Compared to baselines shown in \Cref{fig:exp_hotspots}, \name-DP achieving a 7.8\x to 12.5\x reduction in $P_h$ and a 16.5\x to 27.3\x improvement in minimizing resonator crossing points, underscoring its effectiveness in finetune quantum layouts.

\section{CONCLUSION}
\noindent We introduced \name{}, a quantum legalization and detailed placement framework tailored for robust superconducting quantum processors. It strategically legalizes qubits and resonator separately on substrates, in catering their specific features, then further finetune the layout with detailed placer to enhance the overall robustness of processors.

\bibliographystyle{IEEEtran}
\bibliography{references}

\begin{thebibliography}{10}
\providecommand{\url}[1]{#1}
\csname url@samestyle\endcsname
\providecommand{\newblock}{\relax}
\providecommand{\bibinfo}[2]{#2}
\providecommand{\BIBentrySTDinterwordspacing}{\spaceskip=0pt\relax}
\providecommand{\BIBentryALTinterwordstretchfactor}{4}
\providecommand{\BIBentryALTinterwordspacing}{\spaceskip=\fontdimen2\font plus
\BIBentryALTinterwordstretchfactor\fontdimen3\font minus \fontdimen4\font\relax}
\providecommand{\BIBforeignlanguage}[2]{{%
\expandafter\ifx\csname l@#1\endcsname\relax
\typeout{** WARNING: IEEEtran.bst: No hyphenation pattern has been}%
\typeout{** loaded for the language `#1'. Using the pattern for}%
\typeout{** the default language instead.}%
\else
\language=\csname l@#1\endcsname
\fi
#2}}
\providecommand{\BIBdecl}{\relax}
\BIBdecl

\bibitem{SC}
J.~Koch \emph{et~al.}, ``Charge-insensitive qubit design derived from the cooper pair box,'' \emph{PRA}, vol.~76, no.~4, 2007.

\bibitem{google}
F.~Arute \emph{et~al.}, ``Quantum supremacy using a programmable superconducting processor,'' \emph{Nature}, vol. 574, no. 7779, 2019.

\bibitem{crosstalk}
P.~Mundada \emph{et~al.}, ``Suppression of qubit crosstalk in a tunable coupling superconducting circuit,'' \emph{PRApplied}, vol.~12, no.~5, 2019.

\bibitem{para_g}
A.~C. Santos, ``Role of parasitic interactions and microwave crosstalk in dispersive control of two superconducting artificial atoms,'' \emph{PRA}, vol. 107, no.~1, 2023.

\bibitem{substrate_limit}
S.~Huang \emph{et~al.}, ``Microwave package design for superconducting quantum processors,'' \emph{PRX Quantum}, vol.~2, no.~2, 2021.

\bibitem{res_crosstalk}
C.~R.~H. McRae \emph{et~al.}, ``Materials loss measurements using superconducting microwave resonators,'' \emph{Review of Scientific Instruments}, vol.~91, no.~9, 2020.

\bibitem{physical_process_2}
R.~Schoelkopf \emph{et~al.}, ``Wiring up quantum systems,'' \emph{Nature}, vol. 451, no. 7179, 2008.

\bibitem{device_crosstalk}
M.~Brink \emph{et~al.}, ``Device challenges for near term superconducting quantum processors: frequency collisions,'' in \emph{IEEE IEDM}.\hskip 1em plus 0.5em minus 0.4em\relax IEEE, 2018.

\bibitem{spurious}
Y.~Y. Gao \emph{et~al.}, ``Practical guide for building superconducting quantum devices,'' \emph{PRX Quantum}, vol.~2, no.~4, 2021.

\bibitem{quantum_progress}
J.~M. Gambetta \emph{et~al.}, ``Building logical qubits in a superconducting quantum computing system,'' \emph{npj quantum information}, vol.~3, no.~1, 2017.

\bibitem{dist_modeling}
P.~Spring \emph{et~al.}, ``Modeling enclosures for large-scale superconducting quantum circuits,'' \emph{PRApplied}, vol.~14, no.~2, 2020.

\bibitem{qplacer}
J.~Zhang \emph{et~al.}, ``Qplacer: Frequency-aware component placement for superconducting quantum computers,'' \emph{arXiv preprint arXiv:2401.17450}, 2024.

\bibitem{dreamplace}
Y.~Lin \emph{et~al.}, ``Dreamplace: Deep learning toolkit-enabled gpu acceleration for modern vlsi placement,'' in \emph{56th Annual DAC}, 2019.

\bibitem{airbridges}
Z.~Chen \emph{et~al.}, ``Fabrication and characterization of aluminum airbridges for superconducting microwave circuits,'' \emph{APL}, vol. 104, no.~5, 2014.

\bibitem{cpw_cross}
H.~Mukai \emph{et~al.}, ``Pseudo-2d superconducting quantum computing circuit for the surface code: proposal and preliminary tests,'' \emph{New Journal of Physics}, vol.~22, no.~4, 2020.

\bibitem{majority}
J.~Zhang, P.~Bogdan, and S.~Nazarian, ``A majority logic synthesis framework for single flux quantum circuits,'' \emph{arXiv preprint arXiv:2301.10695}, 2023.

\bibitem{CSAR}
------, ``C-sar: Sat attack resistant logic locking for rsfq circuits,'' \emph{arXiv preprint arXiv:2301.10216}, 2023.

\bibitem{cQED}
A.~Blais \emph{et~al.}, ``Circuit quantum electrodynamics,'' \emph{Reviews of Modern Physics}, vol.~93, no.~2, 2021.

\bibitem{intro_cQED}
T.~E. Roth \emph{et~al.}, ``An introduction to the transmon qubit for electromagnetic engineers,'' \emph{arXiv preprint arXiv:2106.11352}, 2021.

\bibitem{tunable_coupler}
C.~J. Neill, \emph{A path towards quantum supremacy with superconducting qubits}.\hskip 1em plus 0.5em minus 0.4em\relax University of California, Santa Barbara, 2017.

\bibitem{SC_1}
J.~A. Schreier \emph{et~al.}, ``Suppressing charge noise decoherence in superconducting charge qubits,'' \emph{PRB}, vol.~77, no.~18, 2008.

\bibitem{detune_flux}
A.~Dewes \emph{et~al.}, ``Characterization of a two-transmon processor with individual single-shot qubit readout,'' \emph{PRL}, vol. 108, no.~5, 2012.

\bibitem{rip_gate}
H.~Paik \emph{et~al.}, ``Experimental demonstration of a resonator-induced phase gate in a multiqubit circuit-qed system,'' \emph{PRL}, vol. 117, no.~25, 2016.

\bibitem{multi_layer_problem}
R.~Harris \emph{et~al.}, ``Experimental investigation of an eight-qubit unit cell in a superconducting optimization processor,'' \emph{PRB}, vol.~82, no.~2, 2010.

\bibitem{multi_layer_cpw}
A.~Dunsworth \emph{et~al.}, ``A method for building low loss multi-layer wiring for superconducting microwave devices,'' \emph{APL}, vol. 112, no.~6, 2018.

\bibitem{macro}
X.~Tang \emph{et~al.}, ``Optimal redistribution of white space for wire length minimization,'' in \emph{ASP-DAC}, 2005.

\bibitem{tetris}
T.-C. Chen \emph{et~al.}, ``Ntuplace3: An analytical placer for large-scale mixed-size designs with preplaced blocks and density constraints,'' \emph{IEEE TCAD}, vol.~27, no.~7, 2008.

\bibitem{bin-aided}
H.~Yang, K.~Fung, Y.~Zhao, Y.~Lin, and B.~Yu, ``Mixed-cell-height legalization on cpu-gpu heterogeneous systems,'' in \emph{2022 Design, Automation \& Test in Europe Conference \& Exhibition (DATE)}.\hskip 1em plus 0.5em minus 0.4em\relax IEEE, 2022, pp. 784--789.

\bibitem{abacus}
P.~Spindler \emph{et~al.}, ``Abacus: Fast legalization of standard cell circuits with minimal movement,'' in \emph{ISPD}, 2008.

\bibitem{surface_code}
A.~G. Fowler \emph{et~al.}, ``Surface codes: Towards practical large-scale quantum computation,'' \emph{PRA}, vol.~86, no.~3, 2012.

\bibitem{ibmq}
\BIBentryALTinterwordspacing
IBM, ``Ibm quantum,'' 2023. [Online]. Available: \url{https://quantum-computing.ibm.com/}
\BIBentrySTDinterwordspacing

\bibitem{rigetti}
\BIBentryALTinterwordspacing
Amazon, ``Rigetti superconducting quantum processors,'' 2023. [Online]. Available: \url{https://aws.amazon.com/braket/quantum-computers/rigetti/}
\BIBentrySTDinterwordspacing

\bibitem{xtree}
G.~Li \emph{et~al.}, ``Software-hardware co-optimization for computational chemistry on superconducting quantum processors,'' in \emph{ACM/IEEE 48th Annual ISCA}.\hskip 1em plus 0.5em minus 0.4em\relax ACM/IEEE, 2021.

\bibitem{bv}
E.~Bernstein \emph{et~al.}, ``Quantum complexity theory,'' in \emph{Proceedings of the twenty-fifth annual ACM symposium on Theory of computing}, 1993.

\bibitem{QAOA}
E.~Farhi \emph{et~al.}, ``A quantum approximate optimization algorithm,'' \emph{arXiv preprint arXiv:1411.4028}, 2014.

\bibitem{ising}
R.~Barends \emph{et~al.}, ``Digitized adiabatic quantum computing with a superconducting circuit,'' \emph{Nature}, vol. 534, no. 7606, 2016.

\bibitem{QGAN}
S.~Lloyd \emph{et~al.}, ``Quantum generative adversarial learning,'' \emph{PRL}, vol. 121, no.~4, 2018.

\bibitem{crosstalk_ding}
Y.~Ding \emph{et~al.}, ``Systematic crosstalk mitigation for superconducting qubits via frequency-aware compilation,'' in \emph{53rd Annual IEEE/ACM MICRO}.\hskip 1em plus 0.5em minus 0.4em\relax IEEE, 2020.

\bibitem{gate_time1}
R.~Barends \emph{et~al.}, ``Diabatic gates for frequency-tunable superconducting qubits,'' \emph{PRL}, vol. 123, no.~21, 2019.

\bibitem{AWR}
\BIBentryALTinterwordspacing
Cadence, ``Awr microwave office,'' 2024. [Online]. Available: \url{https://www.cadence.com/en_US/home/tools/system-analysis/rf-microwave-design/awr-microwave-office.html}
\BIBentrySTDinterwordspacing

\end{thebibliography}

\end{document}